\documentclass[review]{elsarticle}

\usepackage{lineno,hyperref}
\usepackage{amsmath,amssymb,amsfonts}
\usepackage{algorithmic}
\usepackage{graphicx,subfigure}
\usepackage{textcomp}
\usepackage{xcolor}
\usepackage{makecell}
\usepackage{multirow,booktabs}
\usepackage{caption}
\usepackage{bm}
\usepackage{url}

\usepackage{array, color}
\def\BibTeX{{\rm B\kern-.05em{\sc i\kern-.025em b}\kern-.08em
		T\kern-.1667em\lower.7ex\hbox{E}\kern-.125emX}}
\modulolinenumbers[5]

\nocite{*}
\begin{document}

\begin{frontmatter}

\title{Efficient Structurally-Strengthened Generative Adversarial Network for MRI Reconstruction}

\author[mymainaddress]{Wenzhong Zhou}
\ead{zhouwenzhong318@163.com}

\author[mymainaddress]{Huiqian Du\corref{mycorrespondingauthor}}
\cortext[mycorrespondingauthor]{Corresponding author}
\ead{duhuiqian@bit.edu.cn}

\author[mymainaddress]{Wenbo Mei}
\ead{wbmei@bit.edu.cn}

\author[mysecondaryaddress]{Liping Fang}
\ead{lpfang@bit.edu.cn}

\address[mymainaddress]{School of Information and Electronics, Beijing Institute of Technology, Beijing 100081, China}
\address[mysecondaryaddress]{School of Mathematics and Statistics, Beijing Institute of Technology, Beijing 100081, China}

\begin{abstract}
Compressed sensing based magnetic resonance imaging (CS-MRI) provides an efficient way to reduce scanning time of MRI. Recently deep learning has been introduced into CS-MRI to further improve the image quality and shorten reconstruction time. In this paper, we propose an efficient structurally strengthened Generative Adversarial Network, termed ESSGAN, for reconstructing MR images from highly under-sampled \textit{k}-space data. ESSGAN consists of a structurally strengthened generator (SG) and a discriminator. In SG, we introduce strengthened connections (SCs) to improve the utilization of the feature maps between the proposed strengthened convolutional autoencoders (SCAEs), where each SCAE is a variant of a typical convolutional autoencoder. In addition, we creatively introduce a residual in residual block (RIRB) to SG. RIRB increases the depth of SG, thus enhances feature expression ability of SG. Moreover, it can give the encoder blocks and the decoder blocks richer texture features. To further reduce artifacts and preserve more image details, we introduce an enhanced structural loss to SG. ESSGAN can provide higher image quality with less model parameters than the state-of-the-art deep learning-based methods at different undersampling rates of different subsampling masks, and reconstruct a 256$\times$256 MR image in tens of milliseconds.
\end{abstract}


\end{frontmatter}


\section{Introduction}
Magnetic Resonance Imaging (MRI) is a widely used imaging modality for clinical diagnosis because of its non-invasiveness and the ability to effectively depict soft tissue changes. However, the inherently long scanning time not only causes motion artifacts and discomfort to patients, but also confines its applications in the time-critical situations like strokes. In order to reduce the acquisition time, a variety of sparse sampling methods have been proposed, including partial Fourier imaging, parallel imaging, the well-known Compressed Sensing based (CS) \cite{CS} methods and so on.

Lustig \textit{et al.} first applied CS \cite{CS} theory to MR image reconstruction by enforcing sparsity of MR images in wavelet transform domain and gradient domain. This CS based method has attracted great attention immediately since it was proposed because it can accurately reconstruct MR images from highly under-sampled $k$-space data. Over the last ten years, a number of CS-MRI methods \cite{NIPS2012_4630}, \cite{han2017}, \cite{zhao2012}, \cite{jin2016}, \cite{Lingala2011} have emerged, mainly focusing on finding: (1) sparsifying transforms or low rank decomposition tools; (2) surrogate functions to enforce sparsity; (3) numerical algorithms to solve optimization problems. The issues encountered by CS-MRI methods are: (1) applying sparsifying transforms or decomposition tools cannot preserve fine-scale details and anatomical structure successfully; (2) the runtime becomes longer as the complexity of the iterative algorithm increases, which turns the reconstruction not-real-time.

Recently, deep learning has made great progress in visual object recognition, object detection, image super-resolution, denoising, and many other domains. ResNet \cite{Resnet} introduced a residual connection which is critical for the training of very deep networks. Besides, Goodfellow \textit{et al.} \cite{GAN} proposed Generative Adversarial Networks (GANs), in which a generator and a discriminator are trained alternately to play adversarial game. It has received huge attention and has been applied in many fields \cite{pixtopix},\cite{cycleGAN},\cite{Photorealistic},\cite{ESRGAN},\cite{SRFeat},\cite{Denoising} due to its outstanding performance. Over the last three years, researchers started to introduce deep learning to the field of CS-MRI reconstruction, where GAN-based approaches have already received impressive results. In \cite{DAGAN}, Yang \textit{et al.} proposed DAGAN for CS-MRI reconstruction, which outperformed most conventional CS-MRI algorithms in terms of reconstruction accuracy and speed. Quan \textit{et al.} \cite{RefineGAN} proposed the RefineGAN, which cascaded two convolutional autoencoders and introduced a cycle data consistency loss for more accurate CS-MRI reconstruction.

Based on previous studies, we propose, in this paper, a novel efficient structurally strengthened Generative Adversarial Network (ESSGAN) with an enhanced structural loss for MRI reconstruction. Compared with the state-of-the-art deep learning-based CS-MRI methods, the proposed ESSGAN can achieve more accurate MRI reconstruction with less model parameters. Our contributions can be mainly summarized as follows:

\begin{enumerate}
	
	\item We introduce strengthened connections (SCs) between two SCAEs to improve the utilization of the feature maps. 
	\item We design a residual in residual block (RIRB), which gives the encoder and decoder blocks richer texture features and allows SG to efficiently trade off between model complexity and performance.
	\item We introduce an enhanced structural loss which combines the advantages of multi-scale structural similarity index (MS-SSIM) and gradient loss to get more texture details  for SG.
	
\end{enumerate}

\par{The rest of this paper is organized as follows. Section II surveys different deep learning-based CS-MRI reconstruction methods and Generative Adversarial Networks. Section III describes the proposed ESSGAN in detail. In section IV, the datasets, masks and training details are introduced. Besides, we show comparison results between our method and the state-of-the-art deep learning-based CS-MRI methods, and analyze the experimental results. Further, we experimentally demonstrate that each proposed novel component plays an important role in improving the performance of the proposed ESSGAN.  In section V, we make a summary of this paper.
}

\section{Related Work}

\subsection{Deep Learning-Based CS-MRI Methods}

Deep learning has been introduced to the field of CS-MRI relatively later than that to other fields. It was not applied to MRI reconstruction until 2016. At the beginning, researchers tried to find ways to combine deep learning with the optimization problems in CS-MRI. Wang \textit{et al.} \cite{wang2016} proposed a two-phase CS-MRI reconstruction method. The neural network was part of classic CS-MRI and was trained to initialize the conventional CS-MRI or to generate images which used as an additional regularization term. Yang \textit{et al.} \cite{ADMMNet} proposed a deep ADMM-Net which unrolls the alternating direction method of multipliers (ADMM) algorithm. Some parameters of the ADMM algorithm are learned in the process of training, such as penalty parameters and shrinkage functions. Schlemper \textit{et al.} \cite{Cascade} introduced the data consistency layer into a cascaded CNN to make the reconstruction faster and more accurate than DLMRI \cite{DL}. Recently, Generative Adversarial Networks (GANs) were first applied to CS-MRI reconstruction called as DAGAN by Yang \textit{et al.} \cite{DAGAN}. The generator of DAGAN used a U-net \cite{Unet} based architecture, the total loss function was composed of a generative adversarial loss, a normalized mean square error (NMSE) loss in image domain, an MSE loss in frequency domain and a perceptual VGG loss. DAGAN improved the quality of reconstructed images dramatically, and outperformed most of classic CS-MRI methods. Quan \textit{et al.} \cite{RefineGAN} proposed the ReconGAN, where a variant of the fully-residual convolutional autoencoder was used as the generator and a cycle data consistency loss was introduced. To further improve reconstruction accuracy, the RefineGAN was proposed, in which the generator was formed by concatenating the generators of ReconGAN. 

The successful applications of deep learning in CS-MRI reconstruction are due to the facts that: (1) MR images lie on or are close to a low dimensional manifold, while deep learning methods can learn MR image manifold from massive amount of data; (2) the end-to-end deep convolutional neural networks mentioned above are capable of learning a mapping between zero-filling (ZF) image manifold and fully-sampled image manifold. In the classic CS-MRI methods, it is very difficult to learn the MR image manifold. However, deep learning makes it possible and can establish a mapping between two image manifolds. Therefore, deep learning-based CS-MRI methods can obtain better reconstruction results than the classic CS-MRI methods.

\subsection{Generative Adversarial Networks}

Generative Adversarial Networks \cite{GAN} are composed of a generator (G) and a discriminator (D), which play a two-player minimax adversarial game. G manages to generate a data distribution to fool D whereas D avoids being fooled. In the CS-MRI reconstruction task, G aims to map the zero-filling image $x_0$ to the fully-reconstructed image $G(x_0)$ to fool D, and D aims to distinguish the real fully-sampled  image from the fake fully-reconstructed image $G(x_0)$. The minimax adversarial game can be formulated as:
\begin{equation}
\centering
\min_{G}\max_{D}L_{GAN} (G,D) = E_{x\sim{P_{data}(x)}}[\log D(x)]+E_{x\sim{P_{zf}(x)}}[\log (1-D(G(x)))],
\end{equation}
where $P_{data}(x)$ represents the distribution of the real data samples and $P_{zf}(x)$ represents the distribution of the zero-filling reconstruction samples. Assume $P_{g}(x)$ is the generated data distribution. If the G is fixed, D can derive the optimal result $D_{op}(x)$:
\begin{equation}
D_{op}(x)=\frac{P_{data}(x)}{P_{data}(x)+P_{g}(x)}.
\end{equation}
When the optimal D is obtained, the minimax adversarial game will turn to determine a generator by solving a minimization problem. It can be reformulated as:
\begin{equation}
\begin{split}
&\min_{G}L_{GAN} (G,D_{op})\\
&=E_{x\sim{P_{data}}}[\log D_{op}(x)]+E_{x\sim{P_{g}}}[\log (1-D_{op}(x))]\\
&=2JSD(P_{data}\parallel{P_{g}})-2\log 2,
\end{split}
\end{equation}
where $ JSD $ denotes the Jensen-Shannon (JS) divergence. The goal of G is to minimize the JS divergence between the real data distribution $P_{data}$ and the generated data distribution $P_g$. The closer $P_g$ is to $P_{data}$, the smaller the JS divergence is between them. This adversarial mechanism can make G generate images that are quite close to the fully-sampled images. 

\section{Method}

\subsection{Problem Formulation}

Let $x\in{C^N}$ denote a fully-sampled MR image, $y\in{C^M}$  represent the measurements in $k$-space where $M<<N$, and $x_0=F^{H}y$ represent the zero-filled reconstruction in which $F$ and superscript $H$ denote the Fourier operator and the Hermitian transpose operation respectively. The aim of CS-MRI reconstruction is to reconstruct $ x $ from $ y $. The relationship between $ y $ and $ x $ can be formulated as:
\begin{equation}
y=Ax+b,
\end{equation}
where $A\in{C^{M\times{N}}}$ represents the under-sampled Fourier operator, $b\in{C^M}$ is the unavoidable noises during the imaging process.

The deep learning-based CS-MRI reconstruction is to train a deep convolutional neural network (CNN) using massive amount of training data to map a ZF image to a fully-reconstructed image, which can be formulated as: 
\begin{align}
\mathop{\arg\min}_{\Theta}L(x,x_{g});\\
x_{g}=f_{CNN}(x_{0},\Theta),
\end{align}
where $x_{g}\in{C^{N}}$ denotes the fully-reconstructed MR image, $f_{CNN}$ is the end-to-end mapping of a CNN, $\Theta$ represents the hidden parameters in the CNN, $L$ is the loss function of the CNN.

\subsection{Total Loss Function}
In the reconstruction scenario, the introduction of GANs aims to generate an image as close as possible to its corresponding fully-sampled image rather than any other real images. However, in the original adversarial process, the adversarial loss allows G to map the input image $x_0$ to any real image in the dataset as long as D cannot distinguish $G(x_0)$ from the real one. In addition, although GANs can preserve some structural and texture features of MRI images in the adversarial process, G may still produce some unrealistic image details or miss some important diagnostic information. Therefore, in order to achieve more accurate MRI reconstruction, an additional and efficient loss function suitable for reconstruction scenario is in need.

\subsubsection{$ L_{2} $ Loss} 
As mentioned in the recent studies \cite{zhao2017} \cite{you2018}, $ L_{2} $  loss has many advantages, such as convexity, differentiability and symmetry. Therefore, $ L_{2} $ loss is usually used as a default loss function. However,  $ L_{2}  $ loss has following shortcomings: (1) $ L_{2}  $ loss is easy to make generated images blurry and unnatural; (2) $ L_{2} $ loss is more inclined to penalize large errors, but it is difficult to penalize small differences, regardless of the structure information of the image; (3)  $  L_{2} $ is easy to stuck in a local minimum. $ L_{2}  $ loss can be expressed as:
\begin{equation}
L_{2}=\dfrac{1}{HW}\left\|G(x_{0})-x\right\|_2^2,
\end{equation}
where \textit{H} and \textit{W} represent the height and the width of a 2D MR image respectively, G denotes the function of the generator.

\subsubsection{$ L_{1} $ Loss} 
Compared to $ L_{2}  $ loss, $ L_{1}  $ loss can avoid some of the shortcomings of $ L_{2}  $ loss: (1)  $ L_{1}  $ loss can suppress image blur; (2) $ L_{1}  $ loss has the same tolerance for large errors and small errors; (3) according to \cite{zhao2017},  $ L_{1}  $ loss can get a lower minimum than $ L_{2}  $ loss. $ L_{1}  $ loss can be formulated as:
\begin{equation}
L_{1}=\dfrac{1}{HW}\left|G(x_{0})-x\right|.
\end{equation}

\subsubsection{Enhanced Structural Loss} 
One goal of MRI reconstruction is to preserve texture details. Therefore, a loss function which can penalize the perceptual and structural difference between the fully-reconstructed images and the fully-sampled images is needed. As revealed in \cite{zhao2017} \cite{you2018}, the structural similarity index (SSIM) \cite{SSIM} and the multi-scale structural similarity index (MS-SSIM) \cite{MS-SSIM} are the two most advanced perceptually motivated metrics, which can be used to generate visually high-quality images.

The SSIM is formulated as follows:
\begin{align}
SSIM(x,x_{g})&=\dfrac{2\mu_{x}\mu_{x_{g}}+C_{1}}{\mu_{x}^{2}+\mu_{x_{g}}^{2}+C_{1}}\cdot\dfrac{2\sigma_{xx_{g}}+C_{2}}{\sigma_{x}^{2}+\sigma_{x_{g}}^{2}+C_{2}},\\
&=l(x,x_{g})*cs(x, x_{g}),
\end{align}
where $ \mu_{x} $, $ \mu_{x_{g}} $, $ \sigma_{x} $, $ \sigma_{x_{g}} $ represent means and standard deviations of the fully-sampled image $ x $ and the fully-reconstructed image $ x_{g} $ respectively, $ \sigma_{xx_{g}} $ denotes the cross-covariance between $ x $ and $ x_{g} $, $ C_{1} $ and $ C_{2} $ are constants. $ l(x,x_{g}) $, $ cs(x, x_{g}) $ are the first term and the second term in Eq. $(9)$.

The MS-SSIM is formulated as follows:
\begin{equation}
MS\verb|-|SSIM(x,x_{g})=l_{M}^\alpha(x,x_{g})\cdot\prod_{j=1}^Mcs_{j}^{\beta_{j}}(x, x_{g}),
\end{equation}
where $ l_{M} $ and $ cs_{j} $ are the terms defined in Eq. (10) at scale $ M $ and $ j $ respectively. According to \cite{MS-SSIM}, MS-SSIM outperforms SSIM under an appropriate parameter settings.
We add a penalty related to MS-SSIM in the loss function of the generator by using the following form:
\begin{equation}
L_{MS\verb|-|SSIM}=1-MS\verb|-|SSIM(x,x_{g}).
\end{equation}

Using structural loss function alone may not preserve texture details well. However, the combination with different structural loss functions which express image details in different ways makes it possible for the generator to maintain image texture features better. For instance, as revealed in \cite{FusionGAN}, the gradients of an image depicting the edges can be used as a loss function to preserve its gradient information, which can be defined as:
\begin{equation}
L_{grad}=\dfrac{1}{HW}\left\|\nabla{x}-\nabla{x_{g}}\right\|_2^2,
\end{equation}
where $ \nabla $ denotes the gradient operator, $ L_{grad} $ represents the gradient loss.

Therefore, we propose an enhanced structural loss $ L_{ES} $ which combines $ L_{MS\verb|-|SSIM} $ and $ L_{grad} $. It can be expressed as:
\begin{equation}
L_{ES}=L_{MS\verb|-|SSIM}+L_{grad}.
\end{equation}
The enhanced structural Loss $ L_{ES} $ can extract richer texture detail information. To the best of our knowledge, this is the first work to combine $ L_{MS\verb|-|SSIM} $ and $ L_{grad} $ for MRI reconstruction.

\subsubsection{Total Loss Function}
Based on the above analysis, the total loss function $L_{total}$ consists of the generative adversarial loss, $ L_{1}  $ loss and the enhanced structural Loss $ L_{ES} $, which can be expressed as:
\begin{equation}
L_{total}=L_{GAN} (G,D)+\alpha L_{1}(G)+\beta L_{ES}(G),
\end{equation}
where $\alpha$ and $\beta$ are the weights that control the balance each other.

\subsection{Overall Architecture of ESSGAN}
The overall architecture of ESSGAN is presented in Fig. 1, which is composed of a structurally strengthened generator (SG) and a discriminator. Note that the shortcuts between the decoder blocks in the first SCAE and the corresponding encoder blocks in the second SCAE are termed as strengthened connections (SCs, the solid red lines in Fig. 1), while the shortcuts between the encoder blocks and the corresponding decoder blocks in each SCAE are termed as typical connections (TCs, the solid blue lines in Fig. 1). SG is a combination of two SCAEs in which the SCs are introduced to improve the utilization of the feature maps between two SCAEs. Besides, in order to further enhance the overall performance of SG, we design a residual in residual block (RIRB) and embed it in SG. In the forward propagation process, the zero-filling reconstruction images are forwardly propagated along the solid black line to generate the fully-reconstructed MR images. The process is expressed as follows.

Assume SCAE $n$ denotes the $n^{th}$ SCAE and $x_{n}$ stands for the output of the SCAE $n$. The feature maps $x_{C(1,n)}^{out}$ are extracted from the $x_{n-1}$, which can be formulated as:
\begin{equation}
x_{C(1,n)}^{out}=C(x_{n-1}),
\end{equation}
where $C$ represents a convolution operator, the subscripts $ C(1,n) $ denotes the function of the first convolution operation (blue circle in Fig. 1) in the SCAE $n$, and the superscripts $ out $ stands for the output of the convolution operator. Then, $x_{C(1,n)}^{out}$  is propagated to $M$ encoder blocks and $M$ decoder blocks in turn. For each encoder block and decoder block, their input and output can be formulated as:
\begin{align}
x_{E(1,n)}^{in}  &= x_{C(1,n)}^{out}+R(x_{C(2,n-1)}^{in}); \\
x_{E(m,n)}^{in}  &= x_{E(m-1,n)}^{out}+R(x_{D(M-m+2,n-1)}^{in}), m=2,3,\cdots,M; \\
x_{E(m,n)}^{out} &= E(x_{E(m,n)}^{in}); \\
x_{D(1,n)}^{in} &= x_{E(M,n)}^{out}+R(x_{D(1,n-1)}^{in}); \\
x_{D(m,n)}^{in} &= x_{D(m-1,n)}^{out}+R(x_{E(M-m+2,n)}^{in}), m=2,3,\cdots,M; \\
x_{D(m,n)}^{out} &= D(x_{D(m,n)}^{in}), 
\end{align}
where Eq. $(17)$ and Eq. $(18)$ denote the input of the encoder block, Eq. $(19)$ denotes the output of the encoder block, Eq. $(20)$ and Eq. $(21)$ denote the input of the decoder block and Eq. $(22)$ is the output of the decoder block. The subscripts $E(m,n)$ and $D(m,n)$ respectively denote the $m^{th}$ encoder block and the $m^{th}$ decoder block in the SCAE $n$. The subscripts $C(2,n)$ represents the function of the second convolution operator (magenta circle in Fig. 1) in the SCAE $n$. The superscripts $in$ and $out$ respectively stand for the input and output of the encoder block, the decoder block or the convolution operator. $E$, $D$ and $R$ respectively represent the functions of the encoder block, the decoder block and the RIRB. Here, one thing that should be paid attention to is that if $n$ equals 1, the  $x_{C(2,n-1)}^{in}$,  $x_{D(M-m+2,n-1)}^{in}$ and $x_{D(1,n-1)}^{in}$ will equal 0 in Eq. $(17)$, Eq. $(18)$ and Eq. $(20)$. The final output $x_n$ of the SCAE $n$ can be expressed as: 
\begin{equation}
x_{n}=x_{n-1}+x_{C(2,n)}^{out}.
\end{equation}
In this paper, the final output of the proposed SG can be formulated as:
\begin{equation}
x_{2}=x_{0}+\sum_{n=1}^{2}x_{C(2,n)}^{out}.
\end{equation}

The discriminator uses the encoding part of the SCAE to distinguish the fully-reconstructed image from the fully-sampled image. In the training process, the discriminator and the generator are trained alternately until a stopping criterion is met.

\begin{figure*}
	\centering
	\includegraphics[scale=0.105]{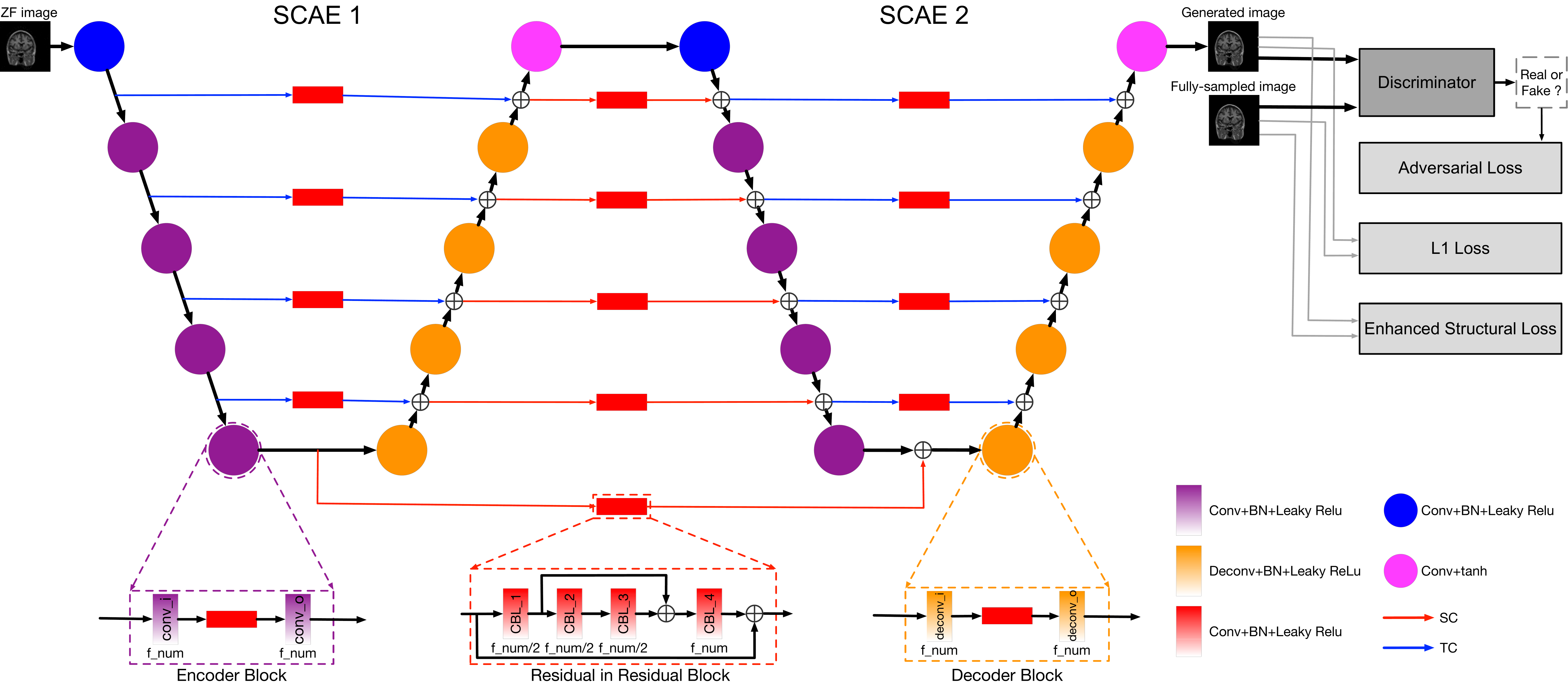}
	\caption{The overall architecture of the proposed ESSGAN: the proposed strengthened generator maps a zero-filling MR image to the reconstructed image to fool the discriminator, however, the discriminator avoids being fooled by distinguishing the fully-sampled image from the reconstructed image.}
	\label{figure1}
\end{figure*}

\subsection{Architecture Details}
In the design of the ESSGAN architecture, we innovatively introduce two important components into the proposed strengthened generator: the SC  and the RIRB. In addition, we introduce a SCAE by embedding the RIRBs to the TCs. Details about each proposed component of ESSGAN will be discussed as follows.

\subsubsection{Strengthened Connection}
Recently, the network performance has been improved by cascading the same CNN architectures (e.g., \cite{RefineGAN}) in a head-to-tail connection style. However, the utilization between the corresponding feature maps of two adjacent CNN architectures is insufficient. To solve this problem, inspired by \cite{LadderNet}, we introduced SCs into the proposed SG. Thus, these SCs dramatically increase the utilization of the corresponding feature maps between two SCAEs. Different from \cite{LadderNet}, the proposed ESSGAN do not share the same weights. To the best of our knowledge, this is the first work to apply SCs to MRI reconstruction. In fact, the SCs can be applied to any other CNN architectures cascaded in a head-to-tail connection style, which is of great significance for improving the utilization of the feature maps. 
\subsubsection{Residual in Residual Block}
As shown in Fig. 1, the residual in residual block (RIRB) is composed of four 2D convolution operators (CBL\_1, CBL\_2, CBL\_3 and CBL\_4), in which each stride of all convolution operators equals 1. From left to right, The filter sizes of the four 2D convolution operators are  $3\times{3}$, $1\times{1}$, $1\times{1}$ and $3\times{3}$ respectively, and the number of filters is $f_{num}/2$, $f_{num}/2$, $f_{num}/2$ and $f_{num}$ respectively. Inspired by \cite{Resnet}, we introduce two residual connections to the RIRB in order to avoid gradient vanishing and thus make training more efficient. Besides, we alternately use $1\times{1}$ filter and $3\times{3}$ filter to exact different texture features. The use of  $1\times{1}$ filter can dramatically reduce the amount of model parameters. Therefore, introducing RIRBs to SG allows it to efficiently trade off between model complexity and performance. The RIRBs are distributed in four places of the ESSGAN architecture, and play a key role in improving network performance. Embedding the RIRBs in the encoder blocks and the decoder blocks can greatly increase the depth of the proposed SG, thus provide higher-level feature maps and increase the expression ability of SG. In addition,  besides the feature information from the previous layer, the encoder or decoder can obtain different feature information from the RIRB by embedding the RIRBs in the TCs and SCs. In other words, the RIRB can give the encoder and decoder blocks richer texture features, thus enhance the feature extraction ability of the proposed SG. In general, the RIRB not only enhances the expression ability of the proposed SG efficiently, but also allows the proposed SG and the discriminator not to suffer from the gradient vanishing problem, making training more efficient. To the best of our knowledge, this is the first work to embed the RIRBs in the shortcuts (TCs and SCs) for MRI reconstruction. More importantly, the method embedding an efficient convolutional block (like RIRB) in the shortcuts can be applied to any CNN architectures, which is of great significance for improving the feature expression ability of the CNN. 
\subsubsection{Strengthened Convolutional AutoEncoder}
We introduced RIRBs into the typical convolutional autoencoder to form the strengthened convolutional autoencoder, dubbed SCAE. Each SCAE contains $M$ encoder blocks (purple circle in Fig. 1), $M$ decoder blocks (orange circle in Fig. 1), two 2D convolution operators (blue and magenta circles respectively in Fig. 1) and the RIRBs. In the encoder block, the conv\_i and conv\_o perform the 2D convolution with the stride 2 and 1 respectively, in which the filter sizes are all $3\times{3}$ and conv\_i performs a down-sampling operation. In the decoder block, the deconv\_i and deconv\_o perform the convolution transpose with stride 1 and 2 respectively, in which the filter sizes are all $3\times{3}$ and deconv\_o can complete an up-sampling operation. The first 2D convolution (blue circle in Fig. 1) is performed with filter size $3\times{3}$, the number of filters is $f_{num}$ and the stride equals 1. The second 2D convolution (magenta circle in Fig. 1) is performed with filter size $3\times{3}$, and the stride equals 1. To match the channel of the input, the number of filters is set to 1.

\begin{figure}
	\centering
	\subfigure[Radial 30\%]{
		\begin{minipage}[b]{0.3\linewidth}
			\includegraphics[width=1.0\linewidth]{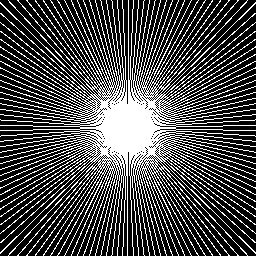}
	\end{minipage}}
	\subfigure[Cartesian 30\%]{
		\begin{minipage}[b]{0.3\linewidth}
			\includegraphics[width=1.0\linewidth]{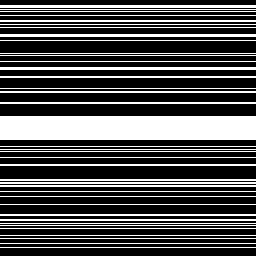}
	\end{minipage}}
	\subfigure[Spiral 30\%]{
		\begin{minipage}[b]{0.3\linewidth}
			\includegraphics[width=1.0\linewidth]{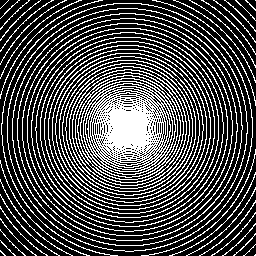}
	\end{minipage}}
	\centering
	\caption{The different subsampling masks with an undersampling rate of 30\%.}
\end{figure}

\section{Experimental Results}

\subsection{Basic Settings}

\subsubsection{Datasets}
We used an MICCAI 2013 grand challenge dataset \footnote{https://my.vanderbilt.edu/masi/workshops/} to train and test all methods. In the training phase, the 100 T1-weighted MRI datasets were selected, which were divided into two groups: 15839 training images (70\%) and 5050 valid images (30\%) including brain tissues. In the testing phase, we randomly selected 50 images out of 10438 testing images to test the performance. It should be noticed that the training images, valid images and testing images are real-valued MR images, which are obtained through preprocessing the complex-valued \textit{k}-space data. Furthermore, they are all independent samples, allowing no overlap among different image sets.

\subsubsection{Masks}
The measurement data $ y $ was obtained by undersampling $k$-space data using three different subsampling masks: radial mask, cartesian mask and spiral mask, each being used for various undersampling rates of 10\%, 20\%, 30\%, and 40\%, corresponding to 10×, 5×, 3.3×, and 2.5× factors of acceleration respectively.

\subsubsection{Comparison Methods and ESSGAN Variants}
Due to the strong performance of the deep learning models, many deep learning-based CS-MRI methods have surpassed most of the classical CS-MRI methods, for example, DAGAN \cite{DAGAN} has better performance than most conventional CS-MRI algorithms (e.g., TV \cite{TV}, SIDWT \cite{SIDWT}, RecPF \cite{RecPF}, DLMRI \cite{DL}, PBDW \cite{PBDW}, PANO \cite{PANO}, Noiselet \cite{Noiselet}, and BM3D \cite{BM3D}). Besides, RefineGAN \cite{RefineGAN} surpasses many deep learning-based CS-MRI methods (e.g., DeepADMM \cite{ADMMNet}, DeepCascade \cite{Cascade}, SingleGAN \cite{Mardani2017}, \cite{yu2017}). Therefore, we just compared the proposed ESSGAN with the state-of-the-art deep learning-based CS-MRI methods. In this study, DAGAN and RefineGAN were chosen to compare with our method in terms of both visual and quantitative quality. Pixel-Frequency-Perceptual-GAN-Refinement (PFPGR) was selected in DAGAN. In addition, in order to ensure the fairness of the comparison, the ZF images were chosen as the inputs for the all methods in the training process. 

To test the effectiveness of the proposed novel components including SC, RIRB and  $L_{ES}$, the following ESSGAN variants were compared: (1) ESSGAN: the full model containing all novel components; (2) ESSGAN-A: the model containing all novel components but SCs; (3) ESSGAN-B: the model containing all novel components but RIRBs embedded in the SCs and TCs; (4) ESSGAN-C: the model containing all novel components but $L_{ES}$.

\subsection{Network Training}
To partly solve the mode collapse problem and enhance network performance, we adopted data augmentation as \cite{DAGAN} for training, such as image flipping, elastic distortion, rotating, shifting, zooming, and brightness adjustment. In the encoder block, the stride is set to 2 to perform a down-sampling operator instead of using an extra max-pooling layer. In the training process, the initial learning rate is 0.0001, and the learning rate is halved every 10 epochs. The Adam optimizer is used with the first-order momentum 0.9 and the second-order momentum 0.999. We adopted the early stopping as the stopping criterion, where the training would stop when the NMSE in the validation set kept increasing 10 times in a row. The proposed ESSGAN was trained by using the following hyperparameters: $M$=4, $\alpha$=200, $\beta$=100, $f_{num}$=64. In addition, we trained and tested the ESSGAN by using tensorflow \footnote{https://www.tensorflow.org} and tensorlayer \footnote{http://tensorlayer.readthedocs.io} on Intel Xeon CPU E5-2630 v4 at 2.2 GHz and a NVIDIA Geforce GTX 1080Ti with 11GB memory. Due to GPU memory constraints, the proposed ESSGAN was trained with a mini-batch size 8. As for the implementations of DAGAN and RefineGAN, we use the source code provided by the authors on GitHub.

\begin{table*}[!htbp]
	\centering
	\renewcommand\tabcolsep{8.7pt} 
	\textbf{Table 1}~~Quantitative results (NMSE / PSNR) of the comparison methods using different undersampling rates of radial subsampling mask.
	\begin{tabular}{ccccc}
		\toprule
		\multirow{2}*{Method} &
		\multicolumn{2}{c}{10\%}   &	\multicolumn{2}{c}{20\%}   \\ 
		&NMSE&PSNR&NMSE& PSNR\\ \hline
		Zero-Filling & 0.246$\pm$0.049 & 30.75$\pm$5.71 & 0.137$\pm$0.040 & 36.12$\pm$6.43   \\ 
		DAGAN   & 0.085$\pm$0.025        & 40.17$\pm$5.34 	   & 0.044$\pm$0.014        & 45.97$\pm$5.47	   \\ 
		RefineGAN   & 0.056$\pm$0.016        & 43.82$\pm$4.58	    & 0.031$\pm$0.015        & 49.17$\pm$3.61  \\ 
		ESSGAN   & {\color{red}0.052$\pm$0.015}        & {\color{red}44.50$\pm$6.20} 	   & {\color{red}0.027$\pm$0.009}        & {\color{red}50.32$\pm$6.44}	       \\
		\bottomrule
	\end{tabular}

	\begin{tabular}{ccccc}
		\toprule
		\multirow{2}*{Method} &
		\multicolumn{2}{c}{30\%} &	\multicolumn{2}{c}{40\%}  \\ 
		&NMSE&PSNR&NMSE& PSNR\\ \hline
		Zero-Filling & 0.084$\pm$0.026 & 40.48$\pm$6.45 & 0.053$\pm$0.016 & 44.47$\pm$6.26 \\ 
		DAGAN        & 0.032$\pm$0.011        & 48.69$\pm$4.88       & 0.027$\pm$0.008       & 50.13$\pm$4.42 \\ 
		RefineGAN  & 0.022$\pm$0.015        & 52.27$\pm$2.87	      & 0.019$\pm$0.015        & 53.83$\pm$2.48 \\ 
		ESSGAN      & {\color{red}0.015$\pm$0.006}        & {\color{red}55.47$\pm$7.04}	        & {\color{red}0.010$\pm$0.004}        & {\color{red}59.55$\pm$7.24}  \\
		\bottomrule
	\end{tabular}
\end{table*}

\begin{table*}[!htbp]
	\centering
	\renewcommand\tabcolsep{8.7pt} 
	\textbf{Table 2}~~Quantitative results (NMSE / PSNR) of the comparison methods at an undersampling rate of 30\% under different subsampling masks.
	\begin{tabular}{ccccc}
		\toprule
		\multirow{2}*{Method} &
		\multicolumn{2}{c}{Cartesian}   &	\multicolumn{2}{c}{Spiral}   \\ 
		&NMSE&PSNR&NMSE& PSNR \\ \hline
		Zero-Filling & 0.188$\pm$0.049 & 33.23$\pm$6.13 &  0.106$\pm$0.030 & 38.26$\pm$6.22  \\ 
		DAGAN   & 0.106$\pm$0.025        & 38.12$\pm$5.67 	        & 0.034$\pm$0.011        & 48.32$\pm$4.98	\\ 
		RefineGAN   & 0.078$\pm$0.021        & 40.81$\pm$4.13	  & 0.023$\pm$0.015        & 51.80$\pm$3.12 \\ 
		ESSGAN   & {\color{red}0.055$\pm$0.013}        & {\color{red}43.89$\pm$5.71} 	   & {\color{red}0.018$\pm$0.007}        & {\color{red}54.53$\pm$7.65}	    \\
		\bottomrule
	\end{tabular}
\end{table*}

\begin{table}[!htbp]
	\centering
	\renewcommand\tabcolsep{22.0pt} 
	\textbf{Table 3}~~Comparison of the number of parameters between different methods.
	\begin{tabular}{cccc}
		\toprule
		Method & DAGAN & RefineGAN & ESSGAN\\
		\midrule
		Params & 146.73M & 156.24M & 35.71M \\
		\bottomrule
	\end{tabular}
\end{table}

\begin{table}[!htbp]
	\centering
	\renewcommand\tabcolsep{8.7pt} 
	\textbf{Table 4}~~Quantitative results (NMSE / PSNR / SSIM) of ESSGAN and its variants at an undersampling rate of 30\% under radial subsampling mask.
	\begin{tabular}{ccccc}
		\toprule
		Method & ESSGAN-A & ESSGAN-B & ESSGAN-C & ESSGAN \\
		\midrule
		NMSE & 0.049$\pm$0.018 & 0.018$\pm$0.006 & 0.016$\pm$0.006 &  {\color{red}0.015$\pm$0.006} \\
		PSNR & 45.07$\pm$4.18 & 53.80$\pm$6.50 & 55.24$\pm$6.98 &  {\color{red}55.47$\pm$7.04} \\
		SSIM & 0.992$\pm$0.004 & {\color{red}0.999$\pm$0.001} & {\color{red}0.999$\pm$0.001} &  {\color{red}0.999$\pm$0.001} \\
		\bottomrule
	\end{tabular}
\end{table}

\begin{table}[!htbp]
	\centering
	\renewcommand\tabcolsep{8.7pt} 
	\textbf{Table 5}~~Quantitative results (NMSE / PSNR / SSIM) of the comparison methods at an undersampling rate of 30\% under radial subsampling mask with additive Gaussian noise ($\mu=0$, $\sigma=1$).
	\begin{tabular}{ccccc}
		\toprule
		Method & Zero-Filling & DAGAN & RefineGAN & ESSGAN \\
		\midrule
		NMSE & 0.093$\pm$0.035 & 0.038$\pm$0.012 & 0.024$\pm$0.015 &  {\color{red}0.023$\pm$0.007} \\
		PSNR & 39.45$\pm$4.43 & 47.25$\pm$4.41 & {\color{red}51.65$\pm$2.99} &  51.54$\pm$4.18 \\
		SSIM & 0.880$\pm$0.057 & 0.997$\pm$0.002 & {\color{red}0.999$\pm$0.001} &  {\color{red}0.999$\pm$0.001} \\
		\bottomrule
	\end{tabular}
\end{table}

\subsection{Result Analysis}
In order to evaluate the reconstructed image quality quantitively, three main image quality metrics are applied: normalized mean-square error (NMSE), peak signal-to-noise ratio (PSNR) and structural similarity (SSIM). Note that NMSE, PSNR and SSIM represent respectively average NMSE, average PSNR and average SSIM in this paper. The subsampling masks are presented in Fig. 2.

\subsubsection{Comprehensive Quantitative Analysis}
Table 1 presents the quantitative results of all methods at different undersampling rates under radial subsampling mask, all results are displayed in the form of mean $\pm$ std (standard deviation). We mark the overall best results in {\color{red}red}. It can be seen that the proposed ESSGAN shows better performance with lower NMSE and higher PSNR than any other comparison methods. In Table 2, the quantitative results are presented at an undersampling rate of 30\% under cartesian and spiral subsampling masks to further test the performance of all these methods. From Table 2 we can observe that significant performance improvement of ESSGAN. It generates better results than DAGAN and RefineGAN under different undersampling patterns. Noticeably, the PSNR values of the proposed ESSGAN are more than 2.7dB higher than those of the most advanced RefineGAN under cartesian and spiral subsampling masks. In addition, the improved performance is not at the expense of a heavier model architecture. In Table 3, in terms of the total number of parameters including the parameters in generator and discriminator, the proposed ESSGAN has 35.71M parameters, which is far less than that of DAGAN and RefineGAN owing mainly to the embedding of the proposed RIRB. In Table 4, we show the quantitative results of ESSGAN variants at an undersampling rate of 30\% under radial subsampling mask. ESSGAN presents better performance than its variants, which illustrates that the removal of any proposed component will affect the performance of the proposed ESSGAN. SCs, RIRBs and $L_{ES}$ play various roles in improving the performance of ESSGAN. In particular, we find that the performance of ESSGAN-A is the worst, which illustrates that SCs are essential for the performance improvement of ESSGAN.

\begin{figure}
	\centering
	\subfigure[]{
		\begin{minipage}[t]{0.48\linewidth}
			\centering
			\includegraphics[width=2.5in]{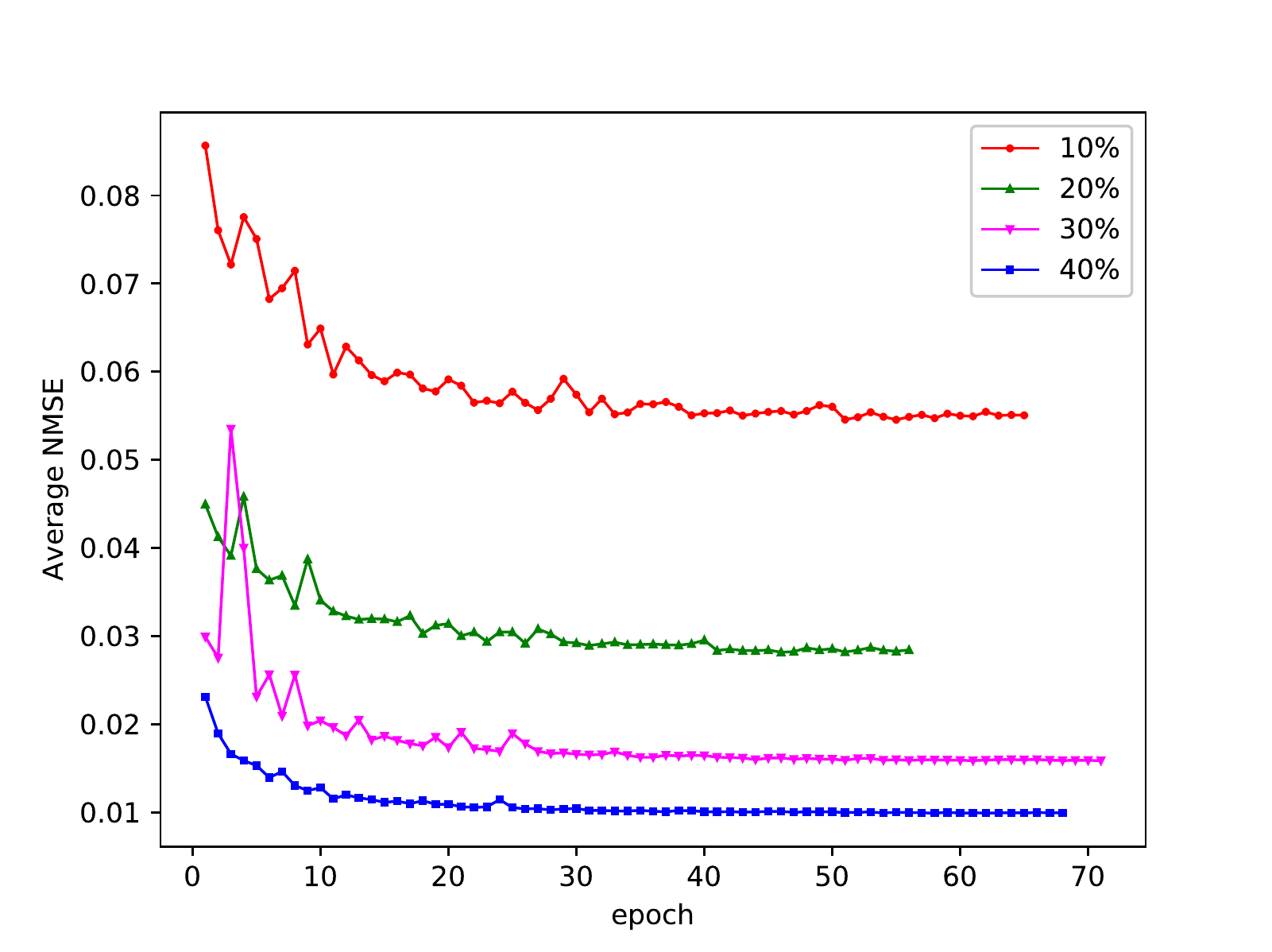}
	\end{minipage}}
	\subfigure[]{
		\begin{minipage}[t]{0.48\linewidth}
			\centering
			\includegraphics[width=2.5in]{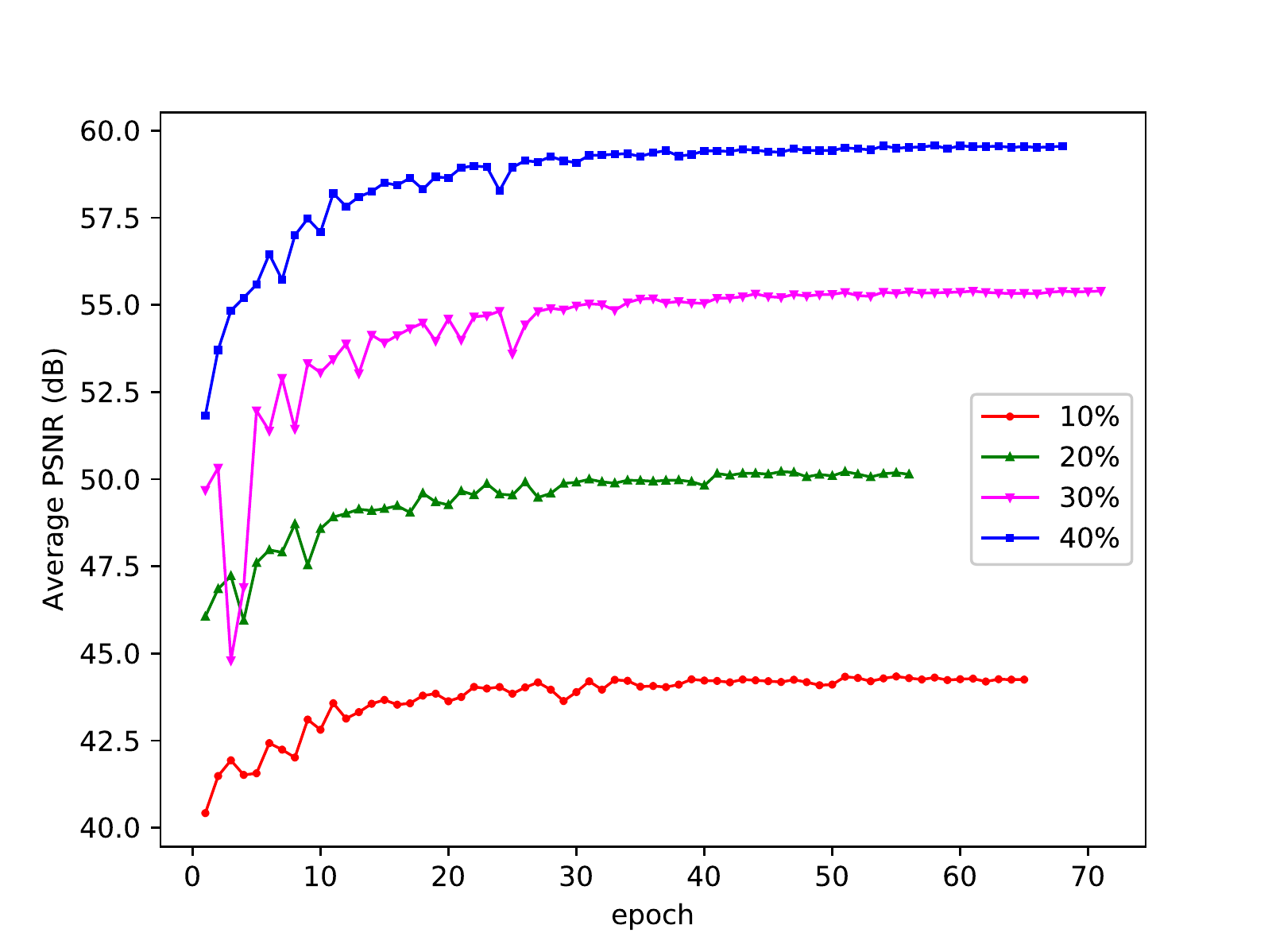}
	\end{minipage}} \\
	\subfigure[]{
		\begin{minipage}[t]{0.48\linewidth}
			\centering
			\includegraphics[width=2.5in]{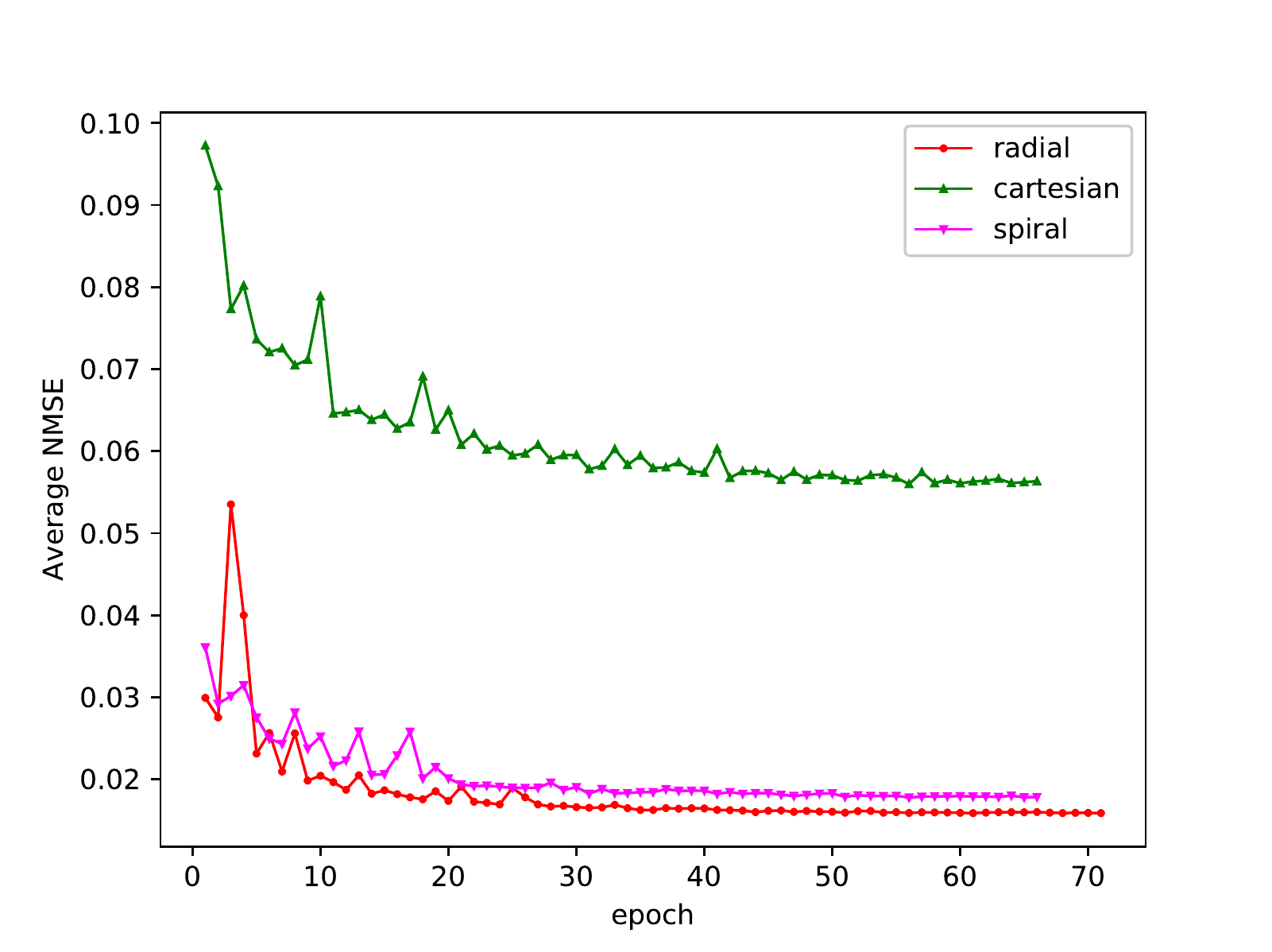}
	\end{minipage}}
	\subfigure[]{
		\begin{minipage}[t]{0.48\linewidth}
			\centering
			\includegraphics[width=2.5in]{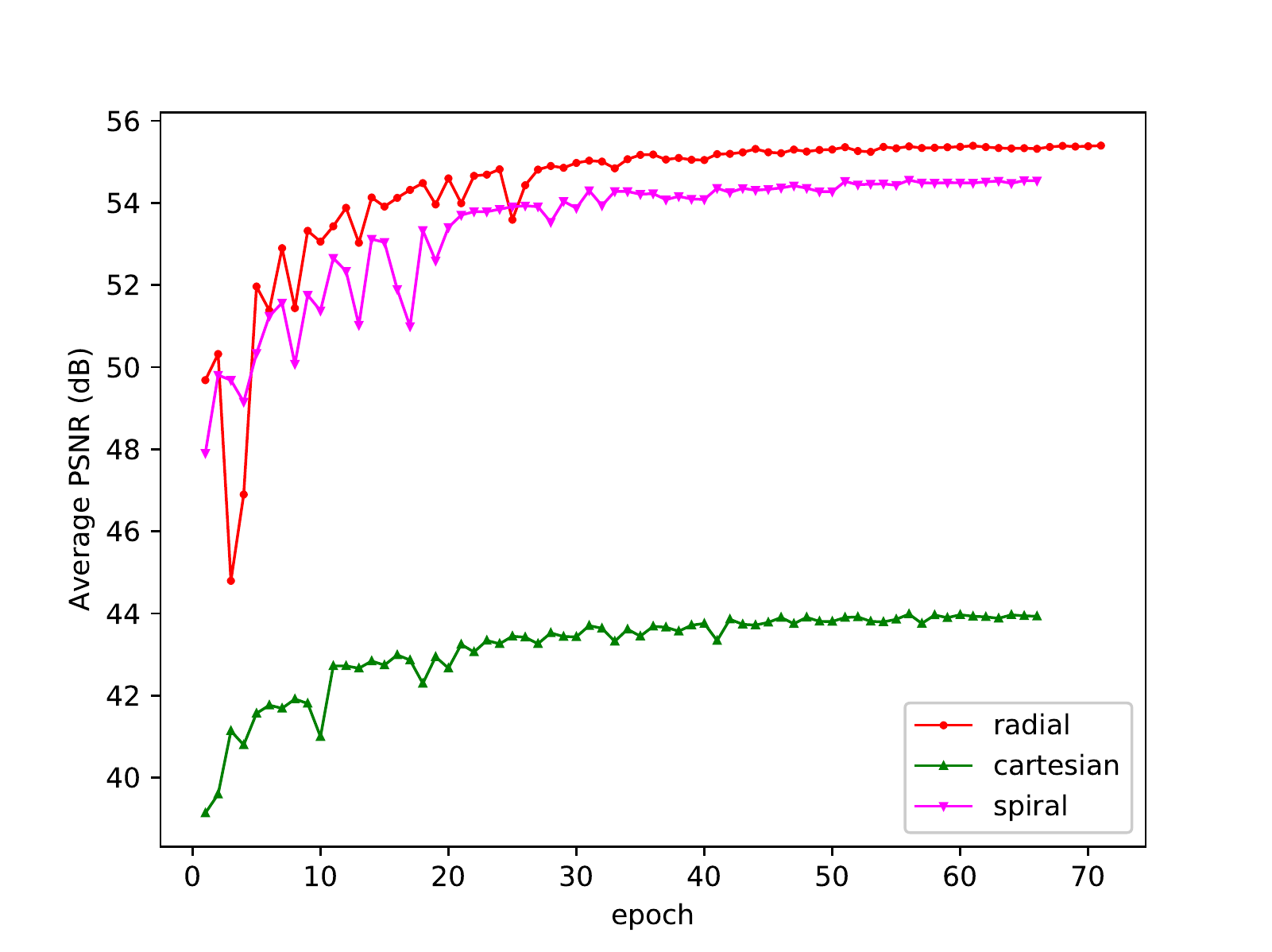}
	\end{minipage}}
	\centering
	\caption{(a) and (b) denote the test results (NMSE / PSNR) of the convergence and stability for ESSGAN at different undersampling rates of the radial mask. (c) and (d) denote the test results (NMSE / PSNR) of the convergence and stability for ESSGAN at an undersampling rate of 30\% under radial, cartesian and spiral subsampling masks.}
\end{figure}

\begin{figure*}
	\centering
	\includegraphics[scale=0.85]{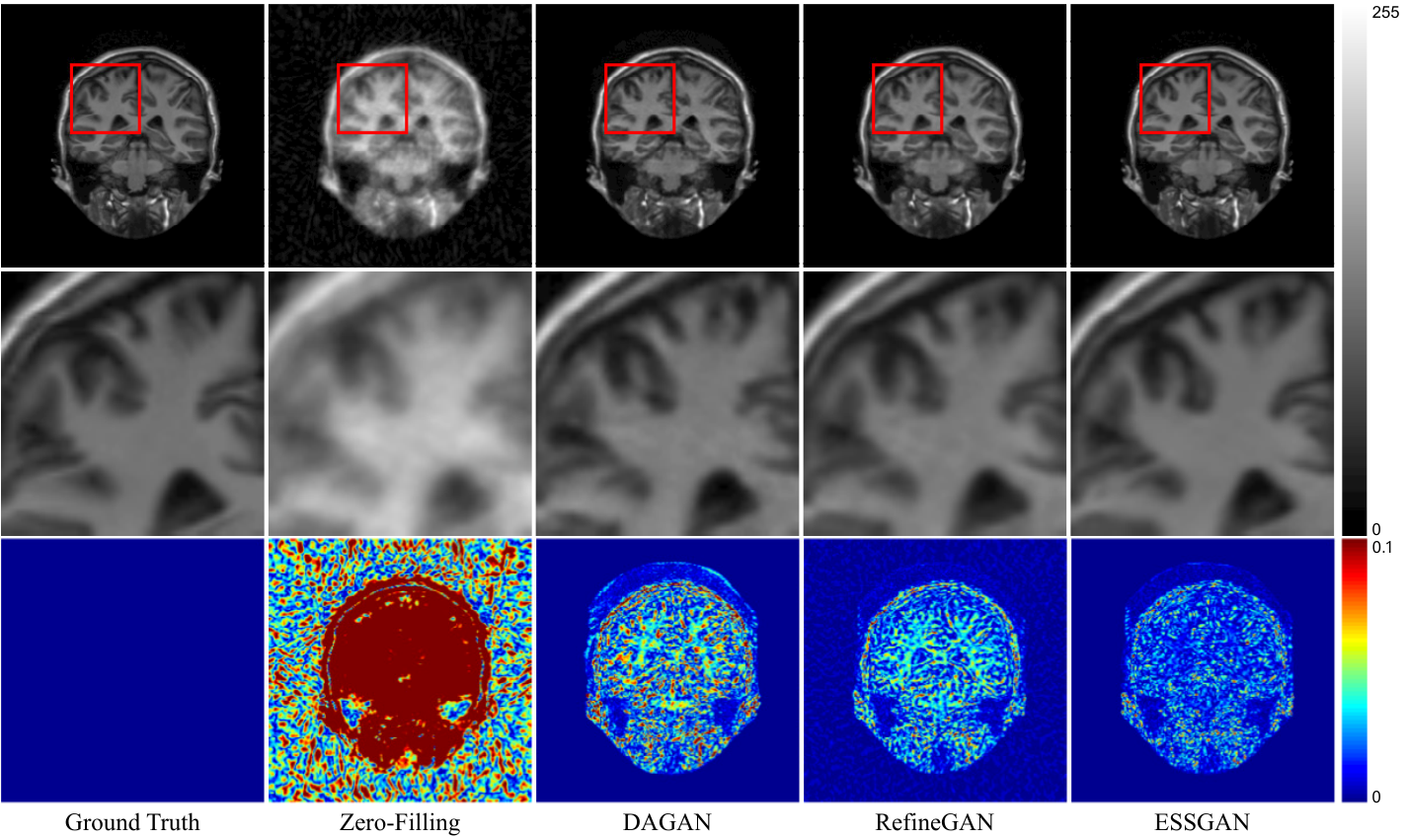}
	\caption{Image quality comparison using radial subsampling mask with an undersampling rate of 10\%: Reconstructed MR images (in the first line), the corresponding zoomed-in ROIs (in the second line) and the corresponding error maps (in the third line).}
	\label{figure_radial_10}
\end{figure*}

\begin{figure*}
	\centering
	\includegraphics[scale=0.85]{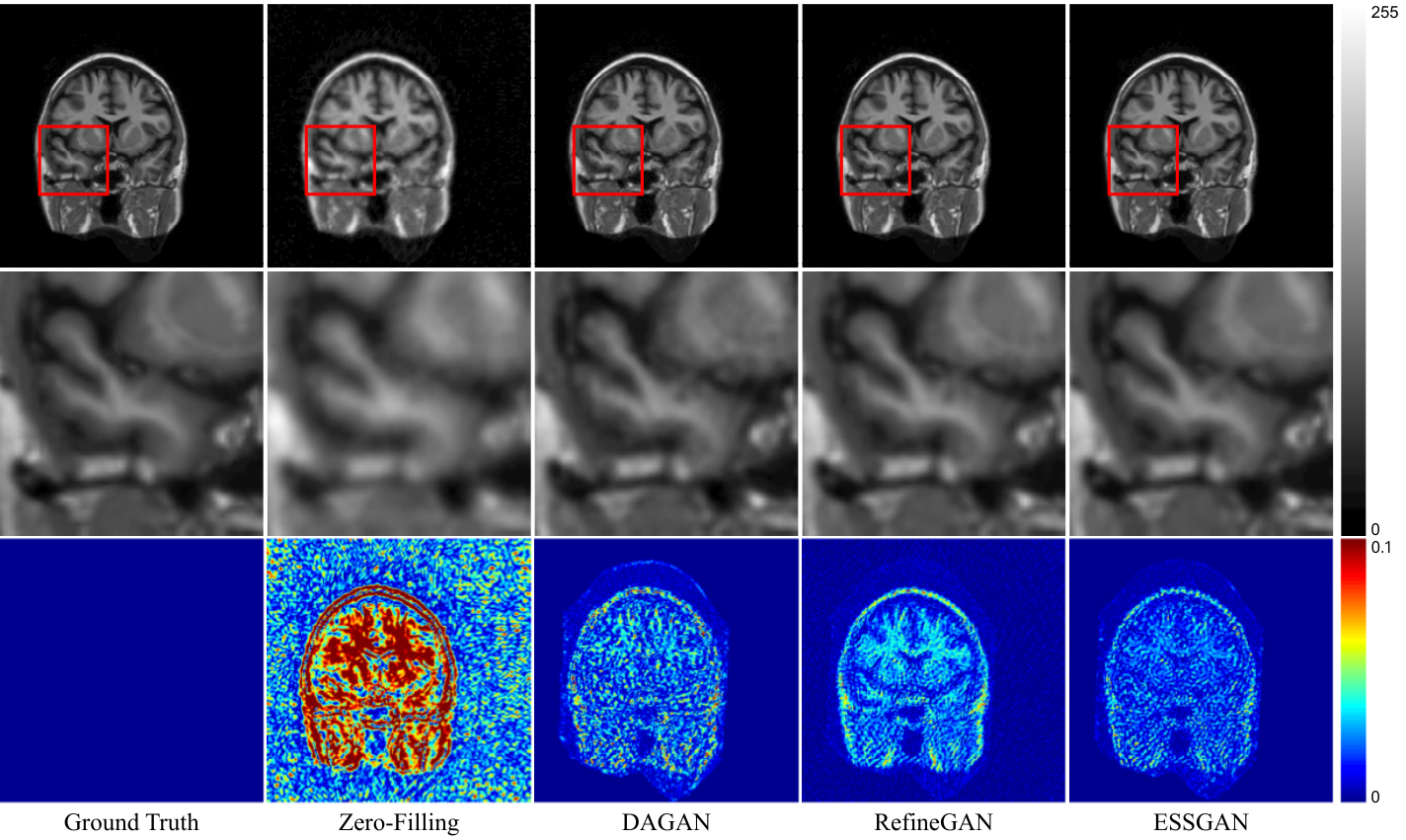}
	\caption{Image quality comparison using radial subsampling mask with an undersampling rate of 20\%: Reconstructed MR images (in the first line), the corresponding zoomed-in ROIs (in the second line) and the corresponding error maps (in the third line).}
	\label{figure_radial_20}
\end{figure*}

\begin{figure*}
	\centering
	\includegraphics[scale=0.85]{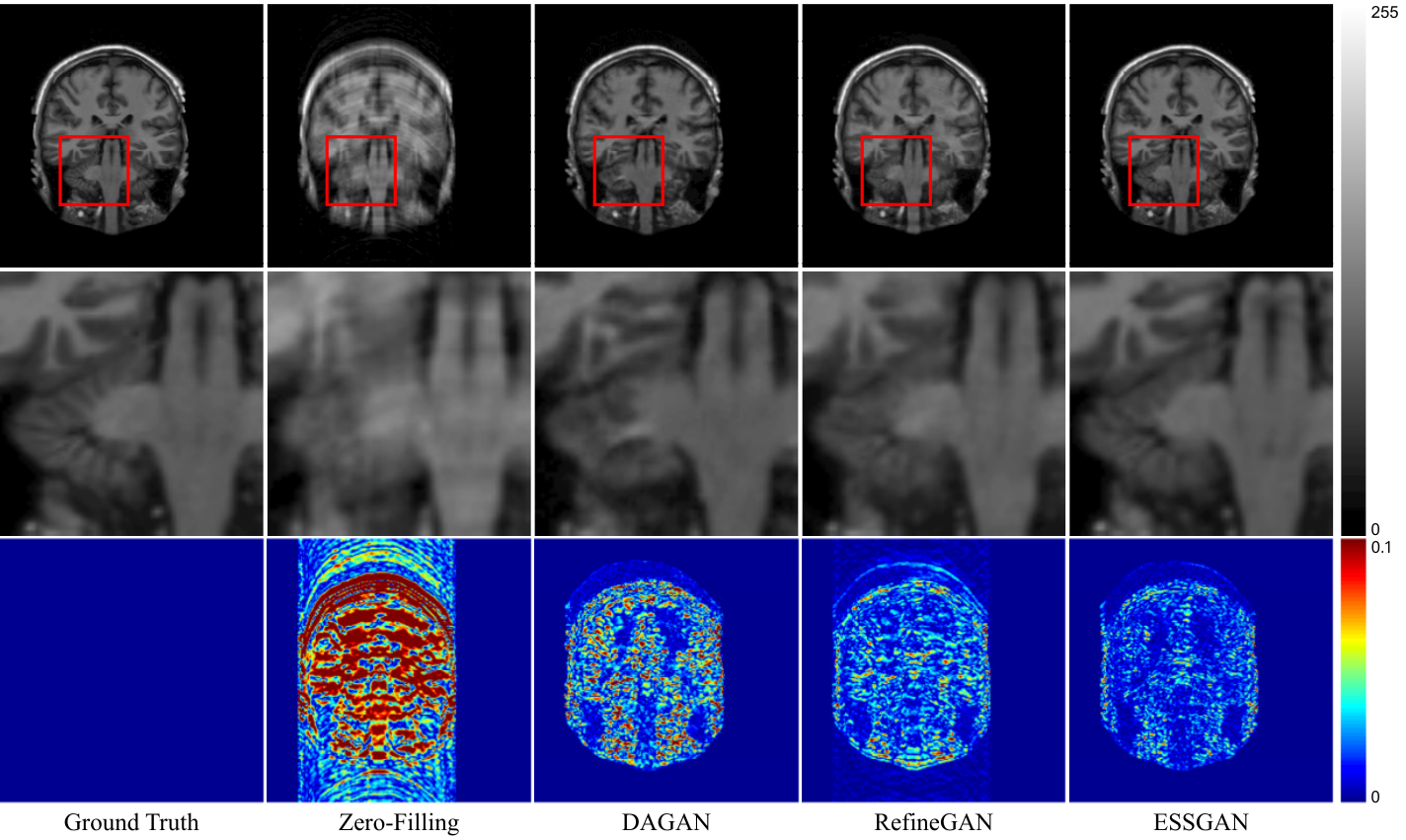}
	\caption{Image quality comparison using cartesian subsampling mask with an undersampling rate of 30\%: Reconstructed MR images (in the first line), the corresponding zoomed-in ROIs (in the second line) and the corresponding error maps (in the third line).}
	\label{figure_cartes_30}
\end{figure*}

\begin{figure*}
	\centering
	\includegraphics[scale=0.85]{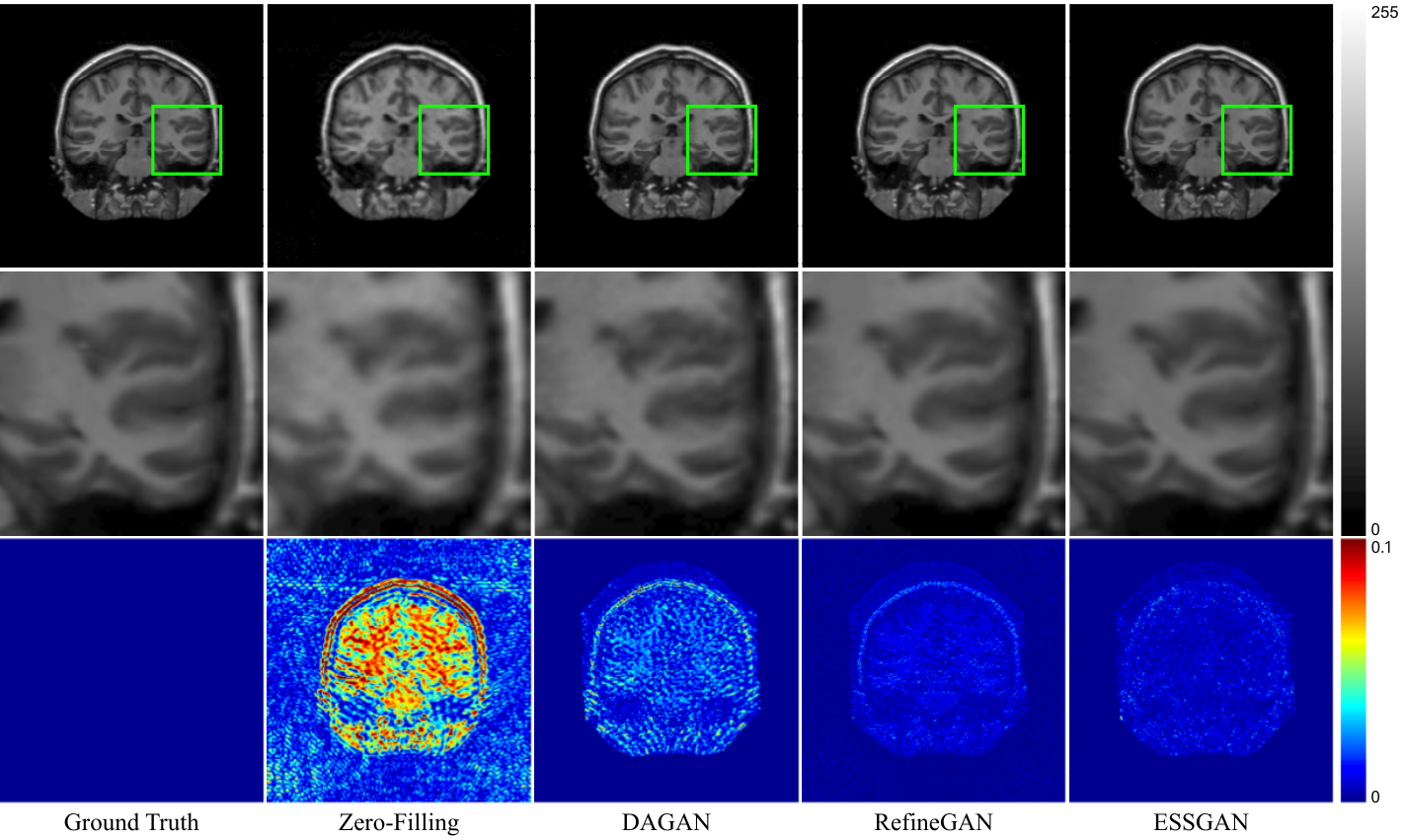}
	\caption{Image quality comparison at an undersampling rate of 30\% of radial subsampling mask with additive Gaussian noise ($\mu=0$, $\sigma=1$): Reconstructed MR images (in the first line), the corresponding zoomed-in ROIs (in the second line) and the corresponding error maps (in the third line).}
	\label{figure_radial_30_an1}
\end{figure*}

\subsubsection{Comprehensive Analysis of ESSGAN in Convergence and Stability}
In the training phase, convergence and stability are very important for the applications of deep learning-based CS-MRI methods. Therefore, we tested the convergence and stability of the proposed ESSGAN. The testing results are presented in Fig. 3. Note that NMSE and PSNR in Fig. 3 are the average NMSE and PSNR in the validation set. (a) and (b) of Fig. 3 respectively present the quantitative results of NMSE and PSNR at different undersampling rates under radial mask. From NMSE and PSNR curves of (a) and (b), we observe that ESSGAN keeps on converging at different undersampling rates under radial mask, and the convergence process is quite stable. (c) and (d) in Fig. 3 respectively present the quantitative results of NMSE and PSNR at an undersampling rate of 30\% under different masks. From (c) and (d) of Fig. 3 we can still find that the NMSE and PSNR curves continually converge to a minimum value steadily.

\subsubsection{Qualitative Visual Analysis}
The qualitative experimental results which demonstrate visual qualities are given in Fig.4-6, which present the reconstructed MR images, the zoomed-in ROIs and the corresponding error maps. Fig. 4 and Fig. 5 illustrate the results from radial subsampling at undersampling rates of 10\% and 20\% respectively. From the zoomed-in ROIs in Fig. 4 and Fig. 5, we observed that ESSGAN can remove more artifacts and reconstruct higher quality images than other comparison methods. From the error maps we see more clearly that the generated MR images reconstructed by the proposed ESSGAN is closer to the ground truth than the comparison methods. Fig. 6 illustrates the results from cartesian subsampling at an undersampling rate of 30\%. Like our findings in Fig. 4 and Fig. 5, we observed from the zoomed-in ROIs in Fig. 6 that the proposed ESSGAN can reconstruct more texture details. More importantly, the texture details reconstructed by the proposed ESSGAN is more realistic and closer to the ground truth. From the error maps we can find that the proposed ESSGAN can achieve more accurate MRI reconstruction with smaller reconstruction errors. 

\subsubsection{Anti-noise Performance Analysis}
In Table 5, the quantitative results are presented with additive Gaussian noise at an undersampling rate of 30\% of radial subsampling mask. Fig. 7 shows the comparison images with additive Gaussian noise. The additive Gaussian noise is chosen with $\mu=0$, $\sigma=1$, where $\mu$ and $\sigma$ represent mean and standard deviation respectively. The additive Gaussian noise is complex-valued because the MRI data in \textit{k}-space is complex-valued. From Table 5 we can find that the quantitative results of ESSGAN and RefineGAN are very close, both of which are better than those of DAGAN. However, from the zoomed-in ROIs of Fig. 7 we can see that the image reconstructed by the proposed ESSGAN is more natural and has less artifacts than that of the most advanced RefineGAN.

In general, the proposed ESSGAN obtains the best results in most cases compared with other comparison methods. Moreover, ESSGAN has lightest network architecture, and in the training phase it has excellent convergence and stability. The reasons for the significant performance improvements are: (1) the introduction of SCs further exploits the advantages of shortcuts, which increases the utilization of the corresponding feature maps; (2) RIRB takes full advantage of the residual connection; (3) Embedding the RIRBs in the encoder blocks, the decoder blocks, the SCs and the TCs enhance the feature expression ability of the proposed SG; (4) $ L_{1}  $ loss and the enhanced structural Loss $ L_{ES} $ help ESSGAN get a lower minimum and preserve more texture details.

\section{Conclusion}
In this paper, we proposed an efficient structurally strengthened Generative Adversarial Network (ESSGAN) with an enhanced structural loss for more accurate MRI reconstruction. The proposed ESSGAN effectively integrates the state-of-the-art deep learning methods: the convolutional autoencoder, the residual network, and GANs. More importantly, we introduced three main innovative components including the SCs, the RIRBs and the enhanced structural loss. The experimental results demonstrated that the proposed ESSGAN can reconstruct more image texture details and remove more artifacts with less model parameters than the state-of-the-art deep learning-based methods.

\section*{}

\end{document}